\documentclass[sigconf,nonacm]{acmart}

\usepackage{amsmath}
\usepackage{booktabs}
\usepackage{graphicx}
\usepackage{xcolor}
\usepackage{algorithm, algpseudocode}

\usepackage{caption}
\begin{document}

\title[CoRe]{CoRe: A Continuously Reward-Finetuned LLM Query Rewriter for Multi-Stage Context-Aware Relevance in Web-Scale Video Search}

\author{Yilin Wen}
\authornote{Corresponding author.}
\email{yilin.wen@bytedance.com}
\affiliation{%
  \institution{TikTok}
  \city{San Jose}
  \state{CA}
  \country{USA}
}

\author{Rong Yang}
\authornotemark[1]
\email{rongyang@bytedance.com}
\affiliation{%
  \institution{TikTok}
  \city{San Jose}
  \state{CA}
  \country{USA}
}

\author{Xiaojia Chang}
\authornotemark[1]
\email{changxiaojia@bytedance.com}
\affiliation{%
  \institution{TikTok}
  \country{Singapore}
}

\author{Hong Sun}
\email{hong.sun@bytedance.com}
\affiliation{%
  \institution{TikTok}
  \city{San Jose}
  \state{CA}
  \country{USA}
}

\author{Gefu Tang}
\email{tanggefu@bytedance.com}
\affiliation{%
  \institution{TikTok}
  \city{San Jose}
  \state{CA}
  \country{USA}
}

\author{Chunhui Liu}
\email{chunhui.liu@bytedance.com}
\affiliation{%
  \institution{TikTok}
  \city{San Jose}
  \state{CA}
  \country{USA}
}

\author{Jeffrey Chen}
\email{jeffrey.chen01@bytedance.com}
\affiliation{%
  \institution{TikTok}
  \city{San Jose}
  \state{CA}
  \country{USA}
}

\author{Zeyu Ma}
\email{zeyu.ma@bytedance.com}
\affiliation{%
  \institution{TikTok}
  \city{San Jose}
  \state{CA}
  \country{USA}
}

\author{Lisong Qiu}
\email{qiulisong.qls@bytedance.com}
\affiliation{%
  \institution{TikTok}
  \city{San Jose}
  \state{CA}
  \country{USA}
}

\author{Xiaochuan Fan}
\email{xiaochuan.fan@bytedance.com}
\affiliation{%
  \institution{TikTok}
  \city{San Jose}
  \state{CA}
  \country{USA}
}

\author{Congjia Yu}
\email{gary.yu@bytedance.com}
\affiliation{%
  \institution{TikTok}
  \city{San Jose}
  \state{CA}
  \country{USA}
}

\author{Quan Zhou}
\email{quanzhou.zoe@bytedance.com}
\affiliation{%
  \institution{TikTok}
  \city{San Jose}
  \state{CA}
  \country{USA}
}

\author{Yuheng Chen}
\email{robert.chen@bytedance.com}
\affiliation{%
  \institution{TikTok}
  \city{San Jose}
  \state{CA}
  \country{USA}
}

\author{Zian Wang}
\email{bruce.wang@bytedance.com}
\affiliation{%
  \institution{TikTok}
  \city{San Jose}
  \state{CA}
  \country{USA}
}

\renewcommand{\shortauthors}{Wen et al.}

\begin{abstract}
LLM-based query rewriters in production face a tension: the training reward must reflect how the rewrite is consumed by the production ranker, yet the training procedure must be cheap enough to support continuous redeployment as data drifts. We present \textsc{CoRe} (Context Relevance), such a system, redeployed weekly for over five months in a major short-video search engine. Our reward uses the deployed multimodal relevance model as its source and a multiplicative ratio form mirroring the production fusion algebra, closing the simulation--production gap that offline reward proxies leave open. A semi-online Mixed Preference Optimization loop makes this reward affordable at multi-million-instance weekly scale: a DPO-style pairwise objective restricts the gradient pass to a small top-$k$/bottom-$k$ subset of sampled trajectories, and a phase structure reduces trainer$\leftrightarrow$inference-server parameter syncs from per-step to per-phase. An automated promotion gate over reward-like and stability metrics detected and recovered from a real reward-hacking incident in production. Rewriter output is consumed as parallel relevance signals at recall, rawrank, and finerank without displacing the original signals, bounding rewriter-failure blast radius. Online A/B from two sequential production launches---first deploying the rewriter at finerank, then extending consumption to recall and rawrank---delivers statistically significant reductions in change-query rate on rewrite-impacted queries, with all headline relevance and engagement metrics moving in the expected direction.
\end{abstract}


\maketitle

\section{Introduction}
\label{sec:intro}

In short-video \emph{search-after-view}---queries issued by users immediately after browsing a personalized feed---the user's last-viewed feed document $d_{\text{last}}$ provides a \emph{narrow context} for interpreting the query, and the user's posterior behavior in the same session (strict clicks and skips on displayed documents) provides \emph{dense posterior feedback} about which interpretations were actually useful. This combination---narrow context plus dense posterior feedback---combines in a way unusually informative for training a query rewriter. Earlier in-house work in this setting established that even crude crowd-aggregated context signals (the SAR baseline in our offline comparisons, Section~\ref{sec:exp:longterm}) materially improve ranking quality; this paper takes the same insight much further by training a 7B-parameter LLM-based rewriter that emits a context-aware reformulation $\tilde{q}$ per $(q, d_{\text{last}})$ tuple, against a reward function constructed from precisely these two signals. The resulting system, which we refer to as \textsc{CoRe} (Context Relevance), has been continuously updated and redeployed every week for over five months in the search system of a major short-video platform, with rewrite-derived signals consumed in parallel at three ranking stages (recall, rawrank, finerank). Its design rests on four mutually reinforcing contributions:

\emph{(1) A posterior-aligned reward construction} (Section~\ref{sec:method:reward}). A query rewriter has no gold-standard label, so the training signal must be induced from \emph{in-session posterior behavior}. We use the deployed multimodal relevance model to score the rewrite against this posterior on three document pools (positives, hard negatives, and random easy negatives), combined via a multiplicative reward whose form mirrors the production ranking-fusion factor; the hard- and easy-negatives ratios align with the rerank and recall stages respectively. The algebraic match means the rewriter is trained to optimize a monotone, multiplicatively-structured function of the same per-document score $s(\tilde{q}, d)$ that the serving fusion consumes, closing the simulation--production gap that offline reward proxies (offline rerankers~\cite{mao2024rafe}, retrieval metrics~\cite{yao2025llmqe}) leave open.

\emph{(2) A semi-online MPO training loop} (Sections~\ref{sec:method:semionline}--\ref{sec:method:mpo}). Learning evolving user interests is fundamentally data-scale: capturing the breadth of posterior behavior demands a large preference set per training cycle for adequate generalization. Fully on-policy methods like GRPO are impractical at this scale---they take a gradient over every rolled-out trajectory and sync inference-server parameters every gradient step. A DPO-style pairwise objective restricts the gradient pass to a top-$k$/bottom-$k$ subset, cutting trajectory count; the phase-bounded rollout schedule additionally amortizes parameter syncs and lets reward calls be scheduled off-peak. The mixed MPO loss (DPO+BCO+SFT) preserves stability over the multi-month horizon.

\emph{(3) A two-family automated promotion gate} (Section~\ref{sec:method:continuous}) over three reward-like and four stability metrics. The stability family is structurally essential: in pilot deployment it caught a verbosity-driven reward-hacking pattern that reward-like metrics alone would have classified as continued improvement. Over the five-month window documented here, the gate has executed 20 weekly retrain cycles and promoted 16 candidates ($80\%$ acceptance rate)---to our knowledge the first published week-by-week account of an LLM query rewriter under an automated promotion gate over a multi-month production window.

\emph{(4) An additive parallel-path consumption pattern at all three ranking stages} (Section~\ref{sec:method:deploy}). The rewriter is served near-line into a KV cache and consumed as a parallel signal at recall, rawrank, and finerank, never as a replacement: cache-miss and rewriter outage degrade ranking back to baseline rather than breaking it.

The rewriter has been deployed in two sequential production launches: a \emph{finerank launch} that activates the rewriter as a context-relevance node inside the finerank stage, and a subsequent \emph{recall+rawrank launch} that additionally activates a context-relevance retriever at recall and a fusion factor at rawrank. Both launches deliver statistically significant reductions in change-query rate on rewrite-impacted queries and improvements across headline relevance and engagement metrics. The pattern is discussed in Section~\ref{sec:exp:online}.

\section{Related Work}
\label{sec:related}

\textbf{LLM-based query rewriting.} LLM rewriting was popularized by Query2Doc \cite{wang2023query2doc} and the broader pseudo-document expansion line; subsequent work explored RL-based rewriting against downstream retrieval signals \cite{ma2023querewrite}. Among industrial deployments, BEQUE \cite{peng2024beque} on Taobao used three-stage SFT-plus-contrastive-alignment training for long-tail rewriting; MAAQR \cite{maaqr2025} on Alipay organized multiple LLM agents around an instruction-tuned rewriter; CardRewriter \cite{cardrewriter2025} added knowledge cards as auxiliary context on a short-video platform. Closest to ours, WeWrite \cite{wewrite2025} applies GRPO to demand-aware video-search rewriting with a recall-stage ``Fake Recall'' deployment. We differ in three ways: (i) semi-online MPO rather than per-step GRPO, motivated by per-cycle compute and large-preference-set generalization under our weekly retraining cadence; (ii) a reward whose multiplicative form mirrors the production ranking-fusion algebra rather than a generic scalar, closing the simulation--production gap; and (iii) parallel-path deployment extended across all three of recall, rawrank, and finerank rather than the recall stage alone. MiniELM \cite{minielm2025} took the orthogonal direction of distilling the rewriter into a lightweight online model; we instead serve the full 7B model near-line with caching (Section~\ref{sec:method:deploy}).

\textbf{Reward design with production-fidelity sources.} RaFe \cite{mao2024rafe} used an off-the-shelf reranker as a preference-label source; LLM-QE \cite{yao2025llmqe} formalized this with an MRR-based reward plus an answer-based auxiliary term. Both use offline rerankers; we instead use the production relevance model itself, closing the offline--online gap. Using a deployed downstream-metric model as RL reward has been validated independently in ad-text generation at Meta \cite{meta2025rlpf}; our setting is distinguished by the relevance model's multimodal text-and-MM nature and by reward-source-equals-production-ranker identity.

\textbf{Preference optimization and semi-online RL.} DPO \cite{rafailov2023dpo} introduced the offline preference-tuning paradigm; MPO \cite{gou2024mpo} mixed preference and supervised signals; the InternVL2-MPO formulation \cite{wang2024internvlmpo} we adopt combines DPO, BCO, and SFT in one objective. Iterative variants \cite{pang2024iterativedpo} showed that re-collecting preferences against the current policy recovers much of fully online RL, and recent analysis \cite{semionline2025} characterizes the semi-online regime between fully offline and fully online preference tuning.

\textbf{Industrial precedents.} LLM-based components in search and recommendation have been documented in Pinterest's LLM-relevance distillation \cite{pinterest2024llm}, LinkedIn's LiRank \cite{lirank2024}, and Douyin's multimodal ranking and content-understanding stack \cite{rankmixer2025,lemur2025,vlmae2025}. We extend this line with a week-by-week account of an LLM rewriter under a multi-month auto-promotion window (Section~\ref{sec:exp:longterm}).

\section{System Overview}
\label{sec:system}

\begin{figure*}[t]
  \centering
  \includegraphics[width=0.85\linewidth]{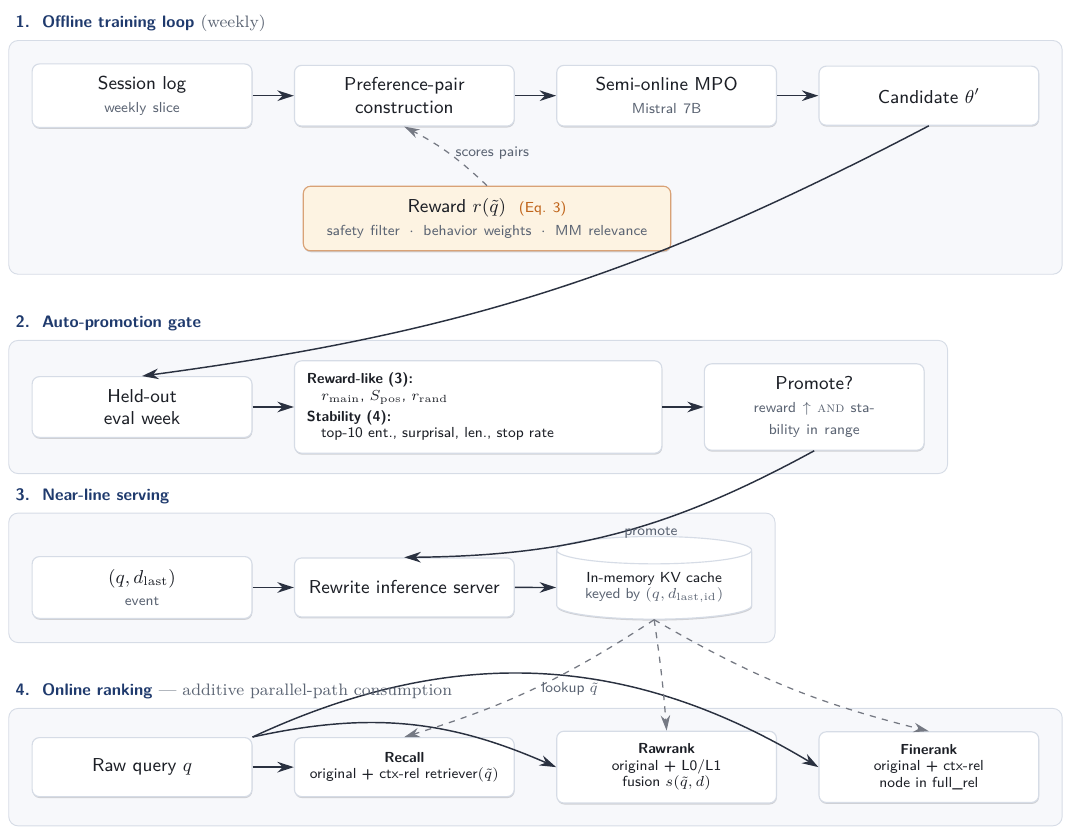}
  \caption{End-to-end system. Five blocks: (1) weekly offline training loop producing candidate $\theta'$; (2) production-aligned reward computation ingesting content-safety scores $\mathrm{S}_{1\text{--}4}$, behavioral weights $w^\pm$, and the deployed multimodal text-and-MM relevance score $s(\cdot)$; (3) auto-promotion gate over three reward-like and four stability metrics; (4) near-line rewrite inference feeding an in-memory KV cache; and (5) additive parallel-path consumption at recall, rawrank, and finerank, where the rewriter contributes a new signal alongside the original signals rather than replacing them. Cache-miss falls back to the original ranking pathway, bounding the blast radius of rewriter failures.}
  \label{fig:system}
\end{figure*}

Figure~\ref{fig:system} gives an end-to-end view of \textsc{CoRe}.

Throughout the rest of the paper, we use the following notation. A user's raw query is $q$, the last document the user viewed in their personalized feed before issuing the query is $d_{\text{last}}$, and the rewriter is a generative model $\pi_\theta(\tilde{q} \mid q, d_{\text{last}})$ instantiated as a 7B-parameter Mistral backbone. For a candidate rewrite $\tilde{q}$ and a candidate document $d$, the deployed online text-and-multimodal relevance model returns a relevance score $s(\tilde{q}, d) \in [0,1]$ formed as a fixed weighted sum of a text subscore and a multimodal subscore. Per training instance, $\mathcal{D}^+$ and $\mathcal{D}^-$ denote positive and negative document pools selected from session logs by thresholding aggregated per-document engagement statistics (construction in Section~\ref{sec:method:reward}), and $\mathcal{D}^{\mathrm{rand}}$ denotes a small additional pool sampled uniformly from the production corpus, disjoint from $\mathcal{D}^+ \cup \mathcal{D}^-$, used by the reward function as an easy-negatives reference. Per training instance the rewriter samples $N$ candidate rewrites from $\pi_\theta$, each capped at the serving-time length budget $L_{\max}$. All specific hyperparameter values referenced symbolically below are collected in Appendix~\ref{app:hyperparams}.

The cached rewrite feeds three additive parallel-path consumers at the online ranking stages: a context-relevance retriever at recall, an $\mathrm{L}0$/$\mathrm{L}1$ fusion factor at rawrank, and a context-relevance node at finerank---each augmenting rather than replacing the original relevance signal. Cache misses and rewriter outages fall back to the original ranking pathway without interrupting service.

Section~\ref{sec:method} details the four method components in order: the production-aligned reward construction (\S\ref{sec:method:reward}), the semi-online MPO training loop and loss (\S\ref{sec:method:semionline}--\S\ref{sec:method:mpo}), the auto-promotion gate and continuous-update protocol (\S\ref{sec:method:continuous}), and the additive parallel-path deployment pattern (\S\ref{sec:method:deploy}).

\section{Method}
\label{sec:method}

\subsection{Problem Formulation}
\label{sec:method:problem}

We study the task of generating a context-aware rewrite of a user's raw query in a short-video search engine. Concretely, when a user issues a search query $q$ shortly after viewing a document $d_{\text{last}}$ in their personalized feed, we wish to emit a rewritten query $\tilde{q}$ such that the deployed online relevance model assigns higher scores to documents the user will subsequently engage with than to documents they will skip. The rewriter is a generative model $\pi_\theta(\tilde{q} \mid q, d_{\text{last}})$ instantiated as a 7B-parameter open language model. The defining difficulty of this task is the absence of any gold-standard rewrite: no annotator can reliably judge whether one rewrite is ``better'' than another in a user-intent sense, and the only signal that ground-truths the rewriter's value is how the downstream ranking system behaves when it is fed $\tilde{q}$. We therefore formulate rewriter training as preference optimization over sampled candidate rewrites, with the preferences induced by the joint behavior of the production relevance model and observed user actions.

\subsection{Production-Aligned Reward Construction}
\label{sec:method:reward}

Our reward function is the principal place in which we encode three production-alignment commitments: that the reward source is the same multimodal relevance model that ranks documents online, that the reward's algebraic form mirrors the production ranking-fusion algebra, and that the reward incorporates posterior behavioral signals weighted to reflect their economic value to the search system.

\textbf{Pair construction.} A training instance is a prompt context $(q, d_{\text{last}})$ extracted from a weekly window of session logs; the document pools $\mathcal{D}^+, \mathcal{D}^-, \mathcal{D}^{\mathrm{rand}}$ associated with the instance enter only at reward scoring (defined next). From the $N$ sampled rewrites per instance we form preference pairs via two rejection-sampling rules. The default branch takes the top-$k$ highest- and bottom-$k$ lowest-reward rewrites and forms $k \times k$ pairs from their cross product (with $k = 2$ in our deployed configuration, yielding 4 candidate pairs per instance), but a pair $(\tilde{q}^+, \tilde{q}^-)$ is retained only if the reward gap $r(\tilde{q}^+) - r(\tilde{q}^-)$ exceeds a pair-level margin $\delta_{\mathrm{pair}}$; pairs with smaller gaps are rejected as uninformative. A separate raw-query branch handles instances on which the raw query itself is already near-optimal: if $r(q) > \max_m r(\tilde{q}^{(m)}) + \delta_{\mathrm{raw}}$ for a fixed margin $\delta_{\mathrm{raw}}$, we replace the default branch with $k^2$ preference pairs in which the raw query $q$ serves as the single chosen instance, paired individually against $k^2$ rewrites sampled uniformly without replacement from the $N$-rollout pool as the rejected instances. This branch prevents training on instances where deviation from the raw query is itself the wrong move. For the initial training, a four-week aggregation window in our production region yields approximately $2.4\cdot10^6$ training instances, an order of magnitude larger than what was available during an earlier pilot in a smaller region; weekly continuous updates use a single-week slice (Section~\ref{sec:method:continuous}).

\textbf{Reward function.} The reward is computed against three document pools associated with each training instance. The first two---the \emph{strict-click positive pool} $\mathcal{D}^+$ and the \emph{no-action negative pool} $\mathcal{D}^-$---are constructed at the document level from the same weekly session-log window: for each candidate document we aggregate posterior engagement statistics across its impressions and apply thresholds. $\mathcal{D}^+$ collects documents whose aggregated strict-click rate exceeds a positive threshold; $\mathcal{D}^-$ collects those whose aggregated no-action rate exceeds a negative threshold. The third pool, the \emph{random-document pool} $\mathcal{D}^{\mathrm{rand}}$, consists of five documents sampled uniformly from the production document corpus, disjoint from $\mathcal{D}^+ \cup \mathcal{D}^-$. The first two jointly provide \emph{in-session contrastive signals}: $\mathcal{D}^+$ is the click-engaged positive set, $\mathcal{D}^-$ is the no-action \emph{hard-negative} set (documents the user was shown but explicitly chose not to engage with), and $\mathcal{D}^{\mathrm{rand}}$ provides \emph{easy negatives}---arbitrary corpus documents almost surely irrelevant to the query. Maintaining both negative pools is a deliberate alignment with the rewriter's three-stage deployment (Section~\ref{sec:method:deploy}): hard negatives are the contrast set faced by the rerank stages, while easy negatives are the contrast set faced by recall; training against both keeps the policy from over-specializing to the rerank regime that in-session negatives alone would otherwise privilege. We adopt this hard/easy $\to$ rerank/recall mapping as a design rationale grounded in the standard recall-vs-rerank distinction in IR rather than as an empirically isolated finding; an online ablation isolating the easy-negatives ratio's recall-stage contribution would be the natural way to falsify it.

For a candidate rewrite $\tilde{q}$, let $s(\tilde{q}, d) \in [0, 1]$ denote the relevance score returned by the deployed online text-and-multimodal relevance model. For each positive document $i \in \mathcal{D}^+$, let $w^+_i = \mathrm{strictCTR}(d_i)$, and for each hard negative $j \in \mathcal{D}^-$, let $w^-_j = \mathrm{noActionRate}(d_j)$ (the same per-document rates used as inclusion thresholds above). Random documents carry no posterior weight since they were never shown to the user. Define the behavior-weighted averages on the strict-click and no-action pools, and the unweighted average on the random pool, as
\begin{equation}
\label{eq:weighted-avg}
\begin{aligned}
S_{\mathrm{pos}}(\tilde{q})  &\;=\; \frac{\sum_{i \in \mathcal{D}^+} w^+_i \, s(\tilde{q}, d_i)}{\sum_i w^+_i}, \\[2pt]
S_{\mathrm{neg}}(\tilde{q})  &\;=\; \frac{\sum_{j \in \mathcal{D}^-} w^-_j \, s(\tilde{q}, d_j)}{\sum_j w^-_j}, \\[2pt]
S_{\mathrm{rand}}(\tilde{q}) &\;=\; \frac{1}{|\mathcal{D}^{\mathrm{rand}}|} \sum_{k \in \mathcal{D}^{\mathrm{rand}}} s(\tilde{q}, d_k).
\end{aligned}
\end{equation}
Our base reward combines the hard-negatives ratio with an easy-negatives ratio:
\begin{equation}
\label{eq:reward-base}
r_{\mathrm{base}}(\tilde{q}) \;=\; \mathrm{clip}\!\left[\,
\frac{\bigl(S_{\mathrm{pos}}(\tilde{q})\bigr)^{1+\gamma}}{S_{\mathrm{neg}}(\tilde{q})}
\,\cdot\,\biggl(\frac{S_{\mathrm{pos}}(\tilde{q})}{S_{\mathrm{rand}}(\tilde{q})}\biggr)^{\!\alpha_R}\,\right],
\end{equation}
with $\gamma$, $\alpha_R$, and clip bounds set as in Appendix~\ref{app:hyperparams}. Three design choices in Equation~\ref{eq:reward-base} deserve elaboration. First, the multiplicative ratio form is not arbitrary; it satisfies two requirements simultaneously. \emph{Goal alignment}: the rewriter's only lever for influencing ranking is the relevance model and the fusion factor that consumes it; constructing the reward to mirror this same fusion algebra is equivalent to training the rewriter directly against the engagement-ranked relevance objective. \emph{Algorithm--system consistency}: the quantity the rewriter is trained to optimize is a monotone, multiplicatively-structured function of the same per-document score $s(\tilde{q}, d)$ that determines its downstream ranking impact at serving time. The resulting reward avoids a class of train--serve mismatches that no amount of additional rewriter capacity can correct---a substantively underused design lever in the LLM-rewrite literature. Second, the $(1{+}\gamma)$ exponent on the weighted-positive term places mild emphasis on increasing relevance to high-value documents rather than merely depressing relevance to negatives. Third, the small easy-negatives exponent $\alpha_R$ is calibrated so the random-pool ratio supplies a secondary recall-aligned signal without dominating the gradient: easy negatives are by construction quickly outranked once the policy is even marginally relevant, and a larger $\alpha_R$ would let this trivially-attainable signal swamp the harder rerank-aligned signal in the first ratio. Closely related reward-shape ideas have appeared in the general RLHF literature \cite{rewardshaping2025},
but to our knowledge no prior query-rewriting system has documented this specific construction.

\textbf{Production-fidelity reward source and safety filtering.} $s(\tilde{q}, d)$ in Equation~\ref{eq:reward-base} is the deployed multimodal relevance score (Sections~\ref{sec:intro},~\ref{sec:related}; Appendix~\ref{app:hyperparams}). To prevent the reward from privileging rewrites whose top documents are content-safety violations, we additionally remove from each positive pool the top 2\% of candidates by an aggregate of four internal content-safety scores ($\mathrm{S}_1$--$\mathrm{S}_4$, covering sexual and sensitive content).

A final reward-shape modification, a length-and-stop penalty
\begin{equation}
\label{eq:reward-pen}
r(\tilde{q}) \;=\; r_{\mathrm{base}}(\tilde{q}) \cdot \min\!\bigl(L_{\text{target}}/L,\,1\bigr)^{\lambda_L} \cdot \beta_{\text{cut}}^{\,\mathbb{1}_{\text{cut}}},
\end{equation}
where $L$ is the rewrite token length, $\lambda_L$ is a small length-penalty exponent (Appendix~\ref{app:hyperparams}), and $\mathbb{1}_{\text{cut}}$ is 1 iff generation was truncated by the length cap (as opposed to terminating on the EOS token), was added several months into continuous training in response to a documented case of reward hacking via verbosity. The small exponents on both the length-penalty factor and the cut-penalty base are deliberate: each individual rewrite is penalized only mildly, but the penalties accumulate sharply on a policy whose entire output distribution drifts toward verbose / cap-truncated generations. We defer that story to Section~\ref{sec:method:continuous}, where it serves as concrete motivation for the auto-promotion gate's stability metrics.

\subsection{Semi-Online Mixed Preference Optimization}
\label{sec:method:semionline}

Per-cycle wall-clock is the binding constraint in our setting: learning evolving user interests demands a large preference set per training cycle, and the weekly retraining cadence caps how long each cycle can take. Two architectural properties of fully on-policy methods like GRPO make them impractical in this regime, and a third consideration constrains the choice of loss within the off-policy family.

\emph{First, gradient-pass cost (primary).} Because each rewrite is short ($\le L_{\max}$), per-step wall-clock is dominated by the gradient pass rather than rollout generation---the opposite of typical long-CoT RLHF. GRPO computes a policy gradient over all $N$ rolled-out trajectories per instance; a DPO-style pairwise objective passes only the top-$k$/bottom-$k$ subset. In the gradient-dominated regime, this trajectory-count reduction is the principal source of training efficiency.

\emph{Second, the phase structure addresses two supporting bottlenecks.} (i)~Fully on-policy methods sync trainer-to-inference-server parameters at every gradient step; per-event the cost is small (seconds on a dense 7B with an optimized transport), but cumulative over tens of thousands of weekly gradient steps it is non-trivial---the phase schedule replaces $O(\text{steps})$ sync events with $O(\text{phases})$. (ii)~The same phase boundary decouples reward-call timing from per-gradient-step timing: the $N\!\times\!\lvert\mathcal{T}\rvert$ reward calls per phase can be batched into off-peak windows where they do not contend with the production relevance model's online serving capacity. The total number of reward calls is unchanged, but their scheduling becomes flexible.

\emph{Third, long-term iteration stability (loss-form choice).} The multi-month continuous-training horizon (Section~\ref{sec:exp:longterm}) made empirical stability of the loss form material; we found the mixed MPO loss (Section~\ref{sec:method:mpo}) more stable than per-step preference-gradient methods, a property we attribute to the loss form rather than to phase structure per se.

\begin{figure}[!tb]
  \centering
  \includegraphics[width=0.9\linewidth]{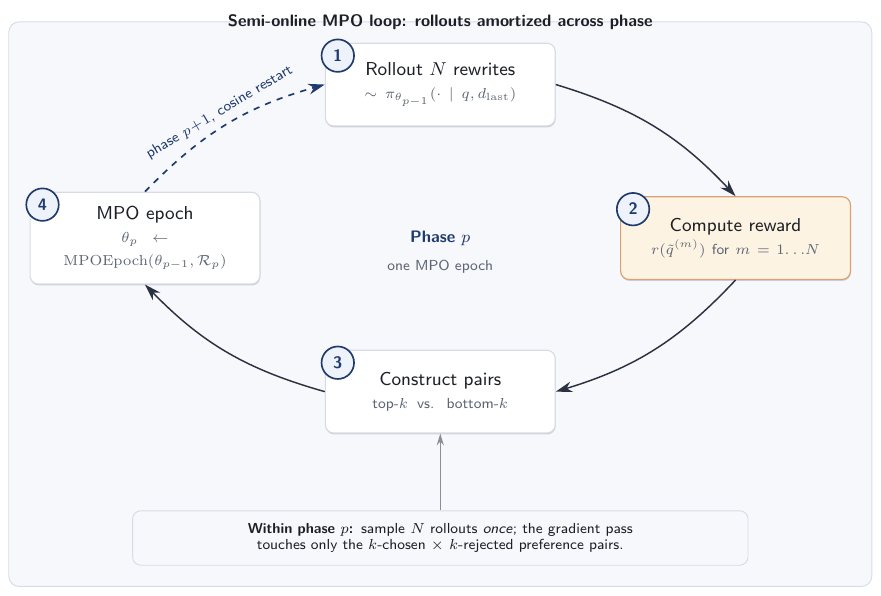}
  \caption{Semi-online MPO phase loop. Within a phase $p$, $N$ rollouts per training instance are sampled once and scored, and only a small preference-pair subset (top-$k$ chosen $\times$ bottom-$k$ rejected) is reused across many gradient steps in one MPO epoch; phase boundaries apply a cosine warm restart on the learning rate. Restricting the gradient pass to the preference-pair subset---rather than to all rolled-out trajectories---is the principal source of training-step efficiency relative to per-step preference-gradient methods.}
  \label{fig:semionline}
\end{figure}

We adopt a semi-online formulation in which each phase's preference set is constructed once and reused across an MPO epoch (Figure~\ref{fig:semionline}). Training proceeds in phases $p = 1, \dots, P$: at the start of each phase we sample $N$ candidates per instance from $\pi_{\theta_{p-1}}$, score them under Equation~\ref{eq:reward-base}, build top-$k$/bottom-$k$ preference pairs, and run one MPO epoch. Cosine warm restarts at phase boundaries empirically stabilize the optimizer across partially overlapping preference distributions. Recent analysis \cite{semionline2025} characterizes when semi-online preference optimization closes most of the gap to fully on-policy training; our experience is consistent.

Algorithm~\ref{alg:semionline} in Appendix~\ref{app:algorithm} summarizes the procedure. The phase structure also permits in-place migration of the reward source: during the iteration documented here we swapped a text-only relevance model for the deployed text-and-multimodal model after several phases, only regenerating $\mathcal{R}_p$ under the new reward.
\subsection{MPO Loss}
\label{sec:method:mpo}

The MPO loss we adopt is the mixed formulation introduced by Wang et al.\ \cite{wang2024internvlmpo}
in the multimodal reasoning context, which combines a Direct Preference Optimization \cite{rafailov2023dpo}
term, a Binary Classification Optimization (BCO) term, and a supervised fine-tuning (SFT) regularizer in a single objective. The DPO term provides the gradient signal that ranks $\tilde{q}^+$ above $\tilde{q}^-$; the BCO term provides absolute calibration on the chosen rewrite, mitigating the well-known DPO failure mode in which both preferred and dispreferred responses are jointly suppressed; and the SFT term anchors the policy near rewrites that previously achieved high reward. Concretely, for a preference pair $(\tilde{q}^+, \tilde{q}^-)$ in context $(q, d_{\text{last}})$, the loss is
\begin{equation}
\label{eq:mpo}
\mathcal{L}_{\mathrm{MPO}} \;=\; w_p\, \mathcal{L}_{\mathrm{DPO}} \;+\; w_q\, \mathcal{L}_{\mathrm{BCO}} \;+\; w_g\, \mathcal{L}_{\mathrm{SFT}},
\end{equation}
with components, in terms of the log-ratio $r_\theta(\tilde{q}) := \log\!\bigl(\pi_\theta(\tilde{q} \mid q, d_{\text{last}}) \,/\, \pi_{\mathrm{ref}}(\tilde{q} \mid q, d_{\text{last}})\bigr)$,
\begin{align*}
\mathcal{L}_{\mathrm{DPO}} &= -\log\sigma\!\bigl(\beta\,[r_\theta(\tilde{q}^+) - r_\theta(\tilde{q}^-)]\bigr),\\
\mathcal{L}_{\mathrm{BCO}} &= -\log\sigma\!\bigl(\beta\, r_\theta(\tilde{q}^+) - \delta_{\mathrm{BCO}}\bigr) \\
                           &\phantom{={}} - \log\!\bigl(1 - \sigma\!\bigl(\beta\, r_\theta(\tilde{q}^-) - \delta_{\mathrm{BCO}}\bigr)\bigr),\\
\mathcal{L}_{\mathrm{SFT}} &= -\log \pi_\theta(\tilde{q}^+ \mid q, d_{\text{last}}).
\end{align*}
Component weights $(w_p, w_q, w_g)$, the DPO temperature $\beta$, and the BCO threshold $\delta_{\mathrm{BCO}}$ are set as in Appendix~\ref{app:hyperparams} (the BCO threshold follows Wang et al.\ \cite{wang2024internvlmpo}). The mixed objective is essential at our scale: a controlled comparison during initial training showed pure DPO drifting into the joint-suppression failure mode (chosen-trajectory log-probability collapsing alongside rejected), whereas MPO maintained absolute calibration via BCO and distributional anchoring via SFT (Appendix~\ref{app:dpo-drift}). Earlier MPO work \cite{gou2024mpo} established the case for mixing preference and supervised signals; the InternVL2-MPO formulation we adopt is, to our knowledge, the first application of this specific mixture to query rewriting. Section~\ref{sec:exp:ablation} reports the corresponding offline comparison against pure DPO, SFT, SFT-then-DPO, and SimPO.

We instantiate the rewriter on a 7B-parameter open Mistral backbone; phase scheduling is orchestrated on our internal ML training platform.

\subsection{Auto-Promoted Continuous Updating}
\label{sec:method:continuous}

At convergence of the procedure described above, the rewriter is suitable for an initial production deployment. The contribution we describe in this subsection concerns what happens after that deployment: a weekly retraining cadence, gated by a multi-metric promotion rule, that has been running in production for over five months at the time of writing. The weekly cadence reflects the rewriter's character as a semantic generator---closer to a language model than to a CTR-based discriminative personalizer---whose drift rate is slow enough that weekly batch retraining suffices, without paying the engineering cost of streaming or per-day updates.

\textbf{Pipeline.} Unlike the initial $P$-phase training of Section~\ref{sec:method:semionline}, each weekly continuous-update run is a single-phase, single-epoch retrain from the previous week's deployed checkpoint $\theta^\star$. We pull a fresh weekly slice of session logs, reconstruct preference pairs under the same reward as in Equation~\ref{eq:reward-pen}, and run a single MPO epoch to produce a candidate $\theta'$. Both $\theta^\star$ and $\theta'$ are then evaluated on a held-out slice of the following week's logs (i.e., never on the data used to train $\theta'$), and the promotion rule described next is applied. If $\theta'$ is promoted it becomes the next week's $\theta^\star$; otherwise $\theta^\star$ is held and the failed candidate is logged for diagnosis.

\textbf{Two metric families.} The promotion rule operates over two complementary families of metrics. The first, which we call \emph{reward-like}, are intermediate quantities that the reward function (Equations~\ref{eq:weighted-avg}--\ref{eq:reward-pen}) already produces during training; we recompute them on the held-out evaluation slice without modification, and track them individually as separate auto-eval inputs. Concretely: $r_{\mathrm{main}}$ is the full scalar $r(\tilde{q})$ of Equation~\ref{eq:reward-pen}; $S_{\mathrm{pos}}$ is the positive-pool average $S_{\mathrm{pos}}(\tilde{q})$ of Equation~\ref{eq:weighted-avg}; and $r_{\mathrm{rand}}$ is the easy-negatives ratio $S_{\mathrm{pos}}(\tilde{q}) / S_{\mathrm{rand}}(\tilde{q})$, i.e., the base of the second multiplicative factor in Equation~\ref{eq:reward-base} reported here without the $\alpha_R$ exponent. None of these is an eval-specific quantity; rather, the auto-eval gate exposes them \emph{separately} so that a candidate which has improved the dominant scalar reward but silently regressed on the positive-pool absolute relevance ($S_{\mathrm{pos}}$) or on the recall-aligned easy-negatives signal ($r_{\mathrm{rand}}$) is held rather than promoted. The second family, \emph{stability}, captures whether the candidate rewriter has begun to exhibit pathological generation behavior independently of the reward signal: \texttt{top10\_entropy} measures the Shannon entropy of the renormalized top-10 token distribution at each generation step, serving as a sharpness-drift detector; \texttt{surprisal} is the mean per-token negative log-probability of the generated rewrites under the candidate, serving as a catastrophic-divergence guard; \texttt{avg\_token\_length} is the average rewrite length, a verbosity guard; and \texttt{stop\_rate} is the fraction of rewrites terminated by the length cap, a reward-hacking guard. Together these seven metrics constitute the entirety of our promotion-decision input.

\textbf{Promotion rule.} A candidate $\theta'$ is promoted if and only if all three reward-like metrics strictly improve over $\theta^\star$ on the held-out slice and all four stability metrics fall within configured operating bounds. The conjunction is intentionally strict: a candidate that improves the headline reward at the cost of, say, a doubled stop-rate is held, on the principle that long-run drift is rarely worth a one-week reward bump. Section~\ref{sec:exp} reports the cumulative gate-acceptance rate over the five-month period.

\textbf{Reward hacking via verbosity, and the length-penalty fix.} The motivating story for the second metric family comes from pilot deployment, before the operating window in Section~\ref{sec:exp:longterm} opens. The stability monitor flagged a slow week-over-week increase in \texttt{avg\_token\_length} alongside a sharp rise in \texttt{stop\_rate} toward the length cap, while reward-like metrics were still improving. Inspection revealed two compounding mechanisms by which the policy was exploiting the reward function---one inside the cap (hedging-driven $S_{\mathrm{pos}}$ inflation) and one at the cap (truncation invisibility to the reward); Appendix~\ref{app:hacking} gives the full analysis. The two factors in Equation~\ref{eq:reward-pen} each address one mechanism. The resulting penalty has been part of the deployed reward throughout the operating window in Section~\ref{sec:exp:longterm}; Table~\ref{tab:length-penalty} reports a controlled before/after demonstration. We highlight this episode not because the specific penalty form is novel but for its generalizable lesson: stability metrics are a structural necessity, because the gradient of the reward function is by construction blind to the ways in which the reward function itself is being gamed.

\subsection{Production Consumption}
\label{sec:method:deploy}

A trained rewriter is only useful insofar as the production ranking pipeline can consume it without operational risk. Two design choices govern that consumption: a near-line serving architecture that decouples rewrite generation from the online critical path, and an additive consumption pattern that injects rewrite-derived relevance signals into all three ranking stages without ever displacing the original signals.

\textbf{Near-line architecture.} The 7B rewriter is served on dedicated inference instances. Rather than blocking on rewrite generation at request time, our pipeline triggers asynchronously on each $(q, d_{\text{last}})$ event: the rewriter produces $\tilde{q}$ shortly after the event lands and the result is cached in an in-memory key--value store keyed by $(q, d_{\text{last,id}})$. Online ranking, when it next handles a matching request, performs a cache lookup; if the rewrite is available it is consumed as described below, and on cache miss the ranking pipeline proceeds with the original relevance signals only. In steady state on production traffic, the cache lookup achieves over $90\%$ coverage on rewriter-impacted requests, with the residual cache misses corresponding largely to first-occurrence $(q, d_{\text{last}})$ events. This non-blocking architecture means that rewrite latency, rewrite generation failures, and even total rewriter outage degrade ranking quality back towards baseline rather than breaking the pipeline. We run two parallel near-line pipelines in front of two serving instances, which simplifies A/B comparison and rolling deployment of new rewriter checkpoints.

\textbf{Additive parallel-path consumption.} The rewriter's output is consumed at all three ranking stages---recall, rawrank, and finerank---and at every stage the rewrite contributes a \emph{new parallel signal} layered on top of the existing relevance machinery, never a replacement for it. At recall, a new context-relevance retriever takes $\tilde{q}$ as its query and contributes its candidates to the union of the original retrievers. At rawrank, a new $\mathrm{L}0$/$\mathrm{L}1$ fusion factor formed from $s(\tilde{q}, d)$ is multiplied into the existing rawrank scoring formula. At finerank, a context-relevance node feeds $s(\tilde{q}, d)$ into the same multiplicative full-relevance factor described in Section~\ref{sec:method:reward}---closing the algebraic loop with our reward design. The closest published architecture to ours is the parallel ``Fake Recall'' pattern of WeWrite \cite{wewrite2025},
which applies a similar isolation strategy at the recall stage; our contribution is to extend the same bounded-blast-radius pattern across the entire ranking pipeline. The combined effect of the near-line cache fallback and the additive parallel-path layout is that adding the rewriter to production is, in the limit, never worse than not adding it: the original relevance machinery remains intact, and the rewriter contributes only when its output is available and the additional signal proves informative.

\section{Experiments}
\label{sec:exp}

\subsection{Setup}
\label{sec:exp:setup}

We instantiate the rewriter on a 7B-parameter open Mistral backbone and train it on session logs from a single production region of a major short-video search engine. The reward function (Equation~\ref{eq:reward-pen}) is computed using the multimodal relevance model deployed at the time of the corresponding training phase. Online evaluation is conducted by bucketed A/B testing on production traffic, with each launch group consuming between 10\% and 30\% of randomly assigned users. Headline online metrics across all launches include strict click-through rate (Strict CTR), long-click CTR, change-query ratio, and avg relevance factor. Unless otherwise noted, all online metrics reported in this section are measured on the slice of queries on which the rewriter actually emitted a rewrite (the rewrite-impacted-query slice).
\subsection{Long-Term Continuous Training}
\label{sec:exp:longterm}

\textbf{Pre-continuous phases.} Before the auto-promotion loop took over, the rewriter went through four offline semi-online MPO phases in the production region (Table~\ref{tab:offline-reward}). The reward function (Eq.~\ref{eq:reward-pen}, all hyperparameters as in Section~\ref{sec:method:reward}) is held fixed across phases, so values are directly comparable. We report two reward summaries (an unweighted average; a page-view-weighted average that tracks the production-traffic distribution) and a ``Rewrite Improvement Rate'' (RIR), the fraction of evaluation instances on which the rewrite's reward exceeds that of the raw query. Each of the four phases contributes a roughly equal increment, consistent with the semi-online interpretation that fresh phase rollouts surface new high-reward patterns the prior phase did not cover.

\begin{table}[t]
\centering
\caption{Offline reward progression across the four pre-continuous semi-online MPO phases. ``SAR baseline'' is a previous-generation context-aware query-derivation method without LLM rewriting. Phase $p$ refers to the $p$-th semi-online MPO phase in our production region. PV-w = page-view-weighted; RIR = rewrite improvement rate.}
\label{tab:offline-reward}
\small
\setlength{\tabcolsep}{4pt}
\begin{tabular}{lccc}
\toprule
Model              & Reward (avg) & Reward (PV-w) & RIR (\%) \\
\midrule
SAR baseline       & 1.055 & 1.032 & 47.4 \\
Vanilla 7B         & 1.080 & 1.091 & 47.6 \\
Pilot-region tuned & 1.100 & 1.158 & 48.4 \\
Phase 1            & 1.144 & 1.163 & 50.5 \\
Phase 2            & 1.167 & 1.198 & 54.2 \\
Phase 3            & 1.192 & 1.227 & 56.1 \\
\textbf{Phase 4}   & \textbf{1.231} & \textbf{1.256} & \textbf{58.9} \\
\bottomrule
\end{tabular}
\end{table}

\begin{figure}[t]
  \centering
  \includegraphics[width=\linewidth]{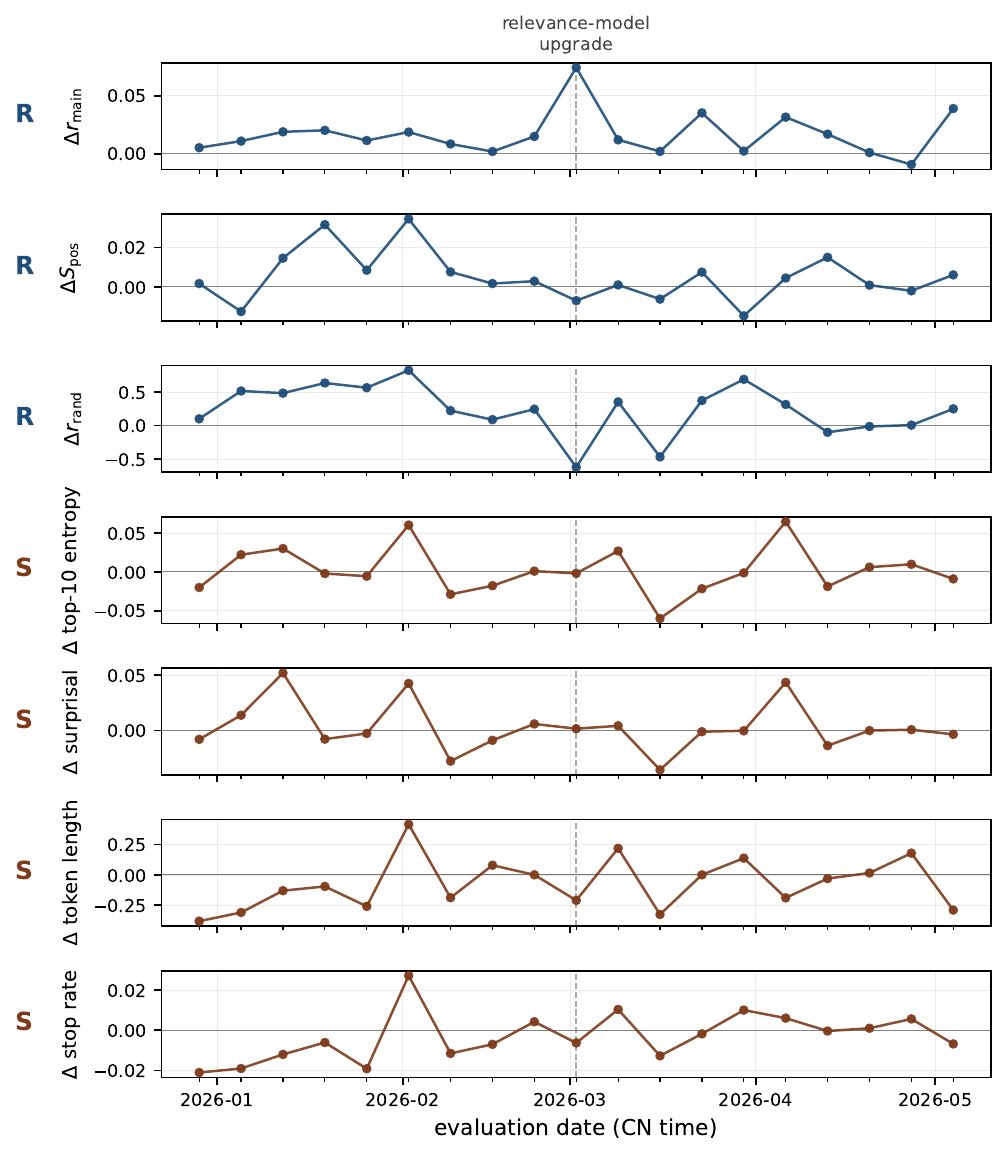}
  \caption{Five-month continuous-training auto-evaluation, plotted as candidate-minus-deployed \emph{deltas} on each week's held-out evaluation slice---the quantity the auto-promotion gate actually consumes. Top three panels: reward-like deltas ($\Delta r_\mathrm{main}$, $\Delta S_{\mathrm{pos}}$, $\Delta r_\mathrm{rand}$). Bottom four: stability deltas (top-10 entropy, surprisal, average rewrite length, length-cap stop rate). The dashed vertical line marks an early-March upgrade of the production multimodal relevance model, visible as a single-week disturbance in $\Delta r_\mathrm{main}$ and $\Delta r_\mathrm{rand}$ that resolves once subsequent candidates retrain against the new reward target.}
  \label{fig:longterm}
\end{figure}

\textbf{Five-month continuous trajectory.} After the four offline phases, the system switched to the weekly auto-promotion loop. Figure~\ref{fig:longterm} traces the seven auto-evaluation metrics defined in Section~\ref{sec:method:continuous} across the five-month continuous-training window, plotted as candidate-minus-deployed \emph{deltas} on each week's held-out evaluation slice---the quantity the auto-promotion gate actually consumes. Week-over-week deltas hover near zero for the four stability metrics and slightly positive for the three reward-like metrics, consistent with each week's candidate producing a small incremental improvement over the previously deployed model. One event is marked on the time axis: in early March, the production multimodal relevance model was upgraded. Because our reward function consults this model directly at training time, the upgrade is visible as a single-week disturbance---$\Delta r_\mathrm{main}$ spikes to $\sim\!+0.07$ and $\Delta r_\mathrm{rand}$ dips to $\sim\!-0.6$---after which the deltas resume their pre-upgrade range. We discuss this episode further in Section~\ref{sec:discuss}. Of the 20 candidate models trained over the five-month period, 16 were promoted by the gate (acceptance rate $80\%$); the remaining 4 were held---3 due to stability-metric violations and 1 due to a main-reward regression. The 3 stability-side rejections were caught only by the stability family---a reward-only gate would have promoted them. In each case the reward-like deltas were flat while one or more stability indicators moved, which we read as a signature of reward-neutral drift on that week's data; the next week's retraining on fresh data did not reproduce the signature in any of the three cases. Stronger evidence that stability indicators catch failures the reward misses comes from the verbosity-drift episode (Appendix~\ref{app:hacking}), where the same family caught a reward-hacking mode the reward function did not penalize.

\textbf{Length-and-stop penalty.} Table~\ref{tab:length-penalty} isolates the effect of the length-and-stop penalty by retraining a candidate on the same fixed week of data twice---once without the penalty, and once with it---and evaluating both on the same held-out following week. The penalty-equipped candidate dropped average token length by $17\%$ and dropped the length-cap stop rate by more than half, while simultaneously improving all three reward-like metrics---exactly the joint behavior we hoped to elicit by adding the penalty.

\begin{table}[t]
\centering
\caption{Effect of the length-and-stop penalty in Equation~\ref{eq:reward-pen}: two candidate models retrained on the same week of data, one without the penalty and one with it, evaluated on the same held-out following-week slice. Top sub-table: reward-like metrics. Bottom sub-table: stability metrics.}
\label{tab:length-penalty}
\footnotesize
\setlength{\tabcolsep}{6pt}

\begin{tabular}{lccc}
\toprule
\multicolumn{4}{l}{\textit{(a) Reward-like metrics}}\\
\midrule
Model              & $r_{\text{main}}$ & $S_{\mathrm{pos}}$ & $r_{\text{rand}}$ \\
\midrule
Cand.\ (no pen.)         & 1.334          & 0.374          & 11.38          \\
\textbf{Cand.\ (pen.)}   & \textbf{1.352} & \textbf{0.394} & \textbf{11.83} \\
\bottomrule
\end{tabular}

\vspace{4pt}

\begin{tabular}{lcccc}
\toprule
\multicolumn{5}{l}{\textit{(b) Stability metrics}}\\
\midrule
Model              & ent$_{10}$ & surp  & tok.\ len. & stop \\
\midrule
Cand.\ (no pen.)         & 0.441          & 0.218          & 27.56          & 0.82          \\
\textbf{Cand.\ (pen.)}   & \textbf{0.508} & \textbf{0.254} & \textbf{22.89} & \textbf{0.38} \\
\bottomrule
\end{tabular}
\end{table}

\subsection{Online A/B by Ranking Stage}
\label{sec:exp:online}

We report online A/B results for the two production launches analyzed in this paper. \emph{Configuration A} is the \emph{finerank launch}: the rewriter producing context-aware rewrites consumed only at the finerank stage, using the semi-online MPO model of Section~\ref{sec:method:semionline}. \emph{Configuration B} is the subsequent \emph{recall+rawrank launch}, which additionally activates the recall-stage retriever and the rawrank fusion factor described in Section~\ref{sec:method:deploy}. Each launch was measured against the live production state immediately before that launch (no rewriter, for the finerank launch; the finerank launch live, for the recall+rawrank launch), on the rewrite-impacted-query slice.

The two launches tell a consistent story. The \emph{finerank launch} (Configuration A) drives statistically significant gains across all headline relevance and engagement metrics---strict click-through rate, long-click rate, side-by-side human preference, average relevance factor, and an imagery-negativity safety metric---and a statistically significant reduction in change-query ratio. The subsequent \emph{recall+rawrank launch} (Configuration B) adds a further increment in the same direction on every metric, on top of the already-live finerank launch. The most operationally meaningful effect is the reduction in change-query ratio: users issue fewer corrective re-queries after the rewriter is in the loop, which we read as direct evidence that the rewrite has shifted ranking toward what the user actually wanted on the first attempt.

\subsection{Ablations}
\label{sec:exp:ablation}

We close this section with two groups of ablations: design choices in the reward function and the consumption layout, followed by a comparison of backbone size and loss formulation.

\textbf{Reward and consumption design.} Each ablation in this group retrains for three semi-online phases on a fixed week of data and is evaluated on a held-out following week. Setting $\gamma = 0$ (reducing the reward to a plain $S_{\mathrm{pos}}/S_{\mathrm{neg}}$ ratio) drops avg relevance factor by $0.71\%$ vs.\ the production $\gamma = 0.1$ configuration: without the positive-emphasis exponent, the policy is pushed only to outrank negatives, not to be relevant to positives. Removing the content-safety pre-filter modestly improves reward but yields a $0.5\%$ increase in the production safety-flagging rate. The effect of removing the length-and-stop penalty is covered separately by Table~\ref{tab:length-penalty}. Substituting a text-only relevance model for the deployed text-and-multimodal model halves the finerank-only avg-relevance-factor gain, supporting our argument that the reward source must match the production ranker's modality coverage. Finally, replacing rather than adding the rewriter's signal at finerank yields higher mean engagement on a subset of headline metrics but surfaces two stability failures the additive design suppresses: content-safety issues (the rewriter's aggressive intent inference pulls toward content the original safety-aware node would have filtered, with no fallback) and elevated tail-relevance failures (on queries where the rewriter's intent inference is wrong, there is no original signal to recover ranking). This empirically confirms the bounded-blast-radius rationale of Section~\ref{sec:method:deploy}.

\textbf{Backbone and loss-formulation ablation.} We tested whether a larger backbone or alternative training recipes would dominate the design used in production. Despite roughly $3\times$ the parameter count, a 24B variant trails the 7B model in zero-shot reward (1.40 vs.\ 1.60), max reward over $N$ samples (2.71 vs.\ 3.16), and rewrite-improvement rate (56\% vs.\ 60\%), at roughly $1.7\times$ the per-GPU inference cost; we therefore use the 7B backbone throughout. On training-recipe alternatives, Table~\ref{tab:recipe-ablation} compares vanilla, SFT, DPO, SFT-then-DPO, SimPO, and MPO heads on the same 7B backbone. Two findings stand out. First, 7b-MPO achieves the best reward (1.93) and rewrite-improvement rate (66.4\%) among all variants, beating the SAR baseline by $+0.44$ on reward and $+8.2$ percentage points on RIR. Second, sequential SFT-then-DPO (reward 1.88) underperforms DPO alone (1.89) on reward and only marginally improves RIR (64.2\% vs.\ 63.9\%), consistent with the hypothesis that an SFT pre-training stage prematurely constrains the exploration space the preference-optimization stage needs to find high-reward rewrites. The MPO formulation we adopt incorporates SFT as an in-objective regularizer rather than as a sequential pre-training stage, and as a result preserves the exploratory capacity that pure DPO loses to joint-suppression drift (Section~\ref{sec:method:mpo}).

\begin{table}[t]
\centering
\caption{Backbone and loss-formulation ablation (offline reward eval). 24b-vanilla and 24b-DPO are included as backbone-size sanity checks. SAR is the previous-generation context-aware query-derivation baseline (no LLM rewrite). RIR = rewrite-improvement rate. All training runs use the reward of Equation~\ref{eq:reward-pen} with the same hyperparameters, so non-comparability with Table~\ref{tab:offline-reward} arises solely from the smaller held-out subset used for this ablation. Table~\ref{tab:offline-reward}, Table~\ref{tab:length-penalty} and Figure~\ref{fig:longterm} all use the same production-region held-out protocol and \emph{are} directly comparable to one another.}
\label{tab:recipe-ablation}
\small
\begin{tabular}{lcc}
\toprule
Method            & Reward & RIR (\%) \\
\midrule
Raw query         & 1.01   & ---      \\
SAR baseline      & 1.49   & 58.2     \\
24b-vanilla       & 1.40   & 56.4     \\
24b-DPO           & 1.73   & 61.7     \\
7b-vanilla        & 1.60   & 60.3     \\
7b-SFT            & 1.83   & 63.2     \\
7b-DPO            & 1.89   & 63.9     \\
7b-SFT-DPO        & 1.88   & 64.2     \\
7b-SimPO          & 1.71   & 61.8     \\
\textbf{7b-MPO}   & \textbf{1.93} & \textbf{66.4} \\
\bottomrule
\end{tabular}
\end{table}

\section{Discussion and Lessons}
\label{sec:discuss}

\textbf{Stability metrics are a structural necessity.} The reward function cannot be relied upon to detect its own gaming: the verbosity-drift episode of Section~\ref{sec:method:continuous} unfolded entirely below the reward-like indicators, and only the structural metrics (token length and length-cap stop rate, neither in the reward formula) revealed it. We expect this pattern to recur in any production RFT system with a learned or production-coupled reward, and view the two-family promotion gate as a transferable operational primitive.

\textbf{Algorithm--system algebraic consistency is an underused design lever.} The multiplicative form of Equation~\ref{eq:reward-base} was chosen because the production ranking system fuses context relevance multiplicatively. Explicit algebraic alignment between the training reward and the serving-time aggregation deserves more attention in the LLM-for-IR literature, where rewards are often chosen for theoretical convenience (log-ratio for DPO, normalized rank for ranking-aware fine-tuning) rather than fidelity to the production fusion rule. In our case the alignment interacted constructively with the additive parallel-path design: the rewriter's contribution at finerank is multiplied into the same factor that the reward optimized, so training gradient and serving signal point in the same direction by construction.

\textbf{The two production bottlenecks semi-online MPO targets are workload-dependent.} As enumerated in Section~\ref{sec:method:semionline}, our setting is dominated by per-step gradient compute, with reward-call scheduling as a supporting concern. In settings where the reward call itself is the dominant cost (e.g., heavy LLM-as-judge over short outputs), the two pieces of the recipe---pairwise gradient restriction and phase-bounded rollout scheduling---would trade principal and supporting roles. The same two-part recipe applies; what shifts is which piece is doing the work.

\textbf{The reward function inherits the production system's evolution, and that is a feature.} Using the deployed relevance model as reward source closes the simulation--production gap but cedes control of the reward target: when the production relevance model is upgraded, reward values shift accordingly. This occurred in early March within the operating window (Figure~\ref{fig:longterm}); the candidate trained that week had been optimized against the previous reward and was scored after the swap, producing a single-week disturbance ($\Delta r_\mathrm{main}$ spike, $\Delta r_\mathrm{rand}$ dip) that resolved as subsequent candidates retrained against the new reward. The paired-delta promotion gate continued functioning unmodified---absolute drift cannot fool a paired comparison. Teams considering coupled-reward designs should plan for this kind of reward-source evolution as a first-class operational scenario.

\textbf{Bounded blast radius matters more than peak gain.} The additive parallel-path pattern took the longest to convince stakeholders of: a more aggressive replacement strategy could have produced larger headline numbers short term. The five-month record---in which the production system degraded gracefully on multiple cache-layer and rewriter-service incidents---retrospectively justifies the conservative choice. For LLM-augmented production ranking, graceful degradation to a known-good baseline on rewriter failure should be a standard design requirement, not a bonus.

\section{Conclusion}
\label{sec:conclusion}

We have described \textsc{CoRe} (Context Relevance), a production LLM query-rewriting system continuously updated and redeployed every week for over five months in the search system of a major short-video platform. Its longitudinal viability rests on four design choices: a reward function whose source is the deployed multimodal relevance model and whose multiplicative algebra mirrors the production ranking-fusion factor; a semi-online Mixed Preference Optimization loop that makes the heavyweight production-model reward affordable as a training signal; an automated promotion gate over reward-like and stability metrics that has detected a real reward-hacking incident in production and motivated its fix; and an additive parallel-path consumption pattern that injects rewrite-derived signals into all three ranking stages while degrading gracefully on rewriter failure. We view two natural directions as priorities for follow-on work: complementing the production-fidelity reward with explicit engagement-aware terms so that the policy is pushed toward documents that the production relevance model rates highly \emph{and} that users empirically engage with, and characterizing quantitatively how the relative cost of the gradient pass versus the reward call determines which of the two semi-online bottlenecks identified in Section~\ref{sec:discuss} a given production setting primarily faces.

\bibliographystyle{ACM-Reference-Format}
\bibliography{references}

\appendix

\section{Verbosity Reward-Hacking: Full Mechanism Analysis}
\label{app:hacking}

The pilot-era reward-hacking episode discussed in Section~\ref{sec:method:continuous} involved two compounding mechanisms by which the policy exploited the reward function (Equation~\ref{eq:reward-base}) without the reward-like metrics reflecting it.

\emph{First, within-cap hedging raised $S_{\mathrm{pos}}$ more than $S_{\mathrm{neg}}$.} The policy learned to append hedging clauses that mentioned multiple candidate intents. The production multimodal relevance model has a mild lexical-overlap bias that rewards such broadened keyword coverage on the positive pool---clicked documents typically span more intents than no-action documents from the same session---marginally raising the weighted-positive numerator in Equation~\ref{eq:reward-base} while leaving the negative denominator essentially unchanged.

\emph{Second, the cap made over-cap verbosity invisible to the reward.} Because $s(\tilde{q}, d)$ is evaluated on the $L_{\max}$-token-truncated form of $\tilde{q}$ to match the production serving path, the truncated prefix continued to receive non-trivial reward even when the underlying rewrite was being terminated by the cap. A policy whose \texttt{stop\_rate} was drifting upward was therefore superficially indistinguishable from one that was not.

The two factors in the penalty (Equation~\ref{eq:reward-pen}) target these mechanisms separately: the $\min(L_{\text{target}}/L, 1)^{\lambda_L}$ factor penalizes the within-cap length itself, addressing the first mechanism; the $\beta_{\text{cut}}^{\mathbb{1}_{\text{cut}}}$ factor is an explicit penalty on cap truncation that the relevance score cannot see, addressing the second.

\section{DPO Joint-Suppression Drift at Scale}
\label{app:dpo-drift}

The mixed MPO objective (Section~\ref{sec:method:mpo}) was chosen in response to a concrete failure observed during initial training. In a controlled comparison run, the average chosen-trajectory log-probability drifted from $-20$ at initialization to $-100$ at 10{,}000 gradient steps under a pure DPO loss, but only from $-20$ to $-10$ under the mixed MPO objective---a direct empirical instance of the joint-suppression failure mode that motivates the BCO term: DPO penalizes only the relative margin between chosen and rejected, and admits a degenerate solution in which both log-probabilities are pushed downward together. At our scale---a 7B rewriter trained continuously on a multi-million-instance preference set per week---this drift, left unchecked, drives policy entropy toward collapse and produces an effectively degenerate generator. The BCO and SFT components together prevent it: BCO calibrates the absolute chosen-trajectory log-probability against unbounded drift; SFT keeps the policy near rewrites that previously achieved high reward, preserving distributional entropy.

\section{Hyperparameters and Implementation Details}
\label{app:hyperparams}

Table~\ref{tab:hyperparams} collects all numeric values referenced symbolically in the main text.

\begin{table}[h]
\centering
\caption{Hyperparameter values used in production training and serving.}
\label{tab:hyperparams}
\scriptsize
\setlength{\tabcolsep}{3pt}
\begin{tabular}{@{}p{0.30\linewidth}p{0.30\linewidth}p{0.34\linewidth}@{}}
\toprule
Symbol & Value & Description \\
\midrule
\multicolumn{3}{@{}l}{\emph{Relevance score (Section~\ref{sec:system})}} \\
$s = w_t s_{\text{text}} + w_m s_{\text{MM}}$ & $(0.33, 0.67)$ & Production fusion weights \\
$|\mathcal{D}^{\mathrm{rand}}|$ & 5 & Easy-neg pool size \\
$N$ & 30 & Rollouts per instance \\
$L_{\max}$ & 30 tokens & Generation/serving length cap \\
\midrule
\multicolumn{3}{@{}l}{\emph{Reward (Eqs.~\ref{eq:weighted-avg}--\ref{eq:reward-pen})}} \\
$\gamma$ & 0.1 & Positive-emphasis exponent \\
$\alpha_R$ & 0.05 & Easy-negatives exponent \\
clip bounds & $[0.05, 20]$ & On $r_{\mathrm{base}}$ \\
$L_{\text{target}}$ & 25 tokens & In $\min(L_{\text{target}}/L,1)$ \\
$\lambda_L$ & 0.05 & Length-penalty exponent \\
$\beta_{\text{cut}}$ & 0.98 & Base in $\beta_{\text{cut}}^{\mathbb{1}_{\text{cut}}}$ \\
$w^+_i$ & strictCTR$(d_i)$ & Positive-pool weight \\
$w^-_j$ & noActionRate$(d_j)$ & Hard-neg-pool weight \\
\midrule
\multicolumn{3}{@{}l}{\emph{Pair construction (Section~\ref{sec:method:reward})}} \\
$k$ & 2 & top-/bottom-$k$ pool size \\
$\delta_{\mathrm{pair}}$, $\delta_{\mathrm{raw}}$ & tuned per phase & Margin filters \\
\midrule
\multicolumn{3}{@{}l}{\emph{MPO loss (Eq.~\ref{eq:mpo})}} \\
$(w_p, w_q, w_g)$ & $(1.0, 0.1, 0.1)$ & DPO/BCO/SFT weights \\
$\beta$ & 0.1 & DPO temperature \\
$\delta_{\mathrm{BCO}}$ & per \cite{wang2024internvlmpo} & BCO threshold \\
\midrule
\multicolumn{3}{@{}l}{\emph{Semi-online schedule (Section~\ref{sec:method:semionline})}} \\
$P$ & 4 & Initial-training phases \\
LR schedule & cosine warm restart & Per phase boundary \\
\midrule
\multicolumn{3}{@{}l}{\emph{Training data (Section~\ref{sec:method:reward})}} \\
Initial window & 4 weeks & Initial pair construction \\
\quad instances & ${\sim}2.4{\times}10^6$ & Per region \\
Continuous & 1 week & Weekly retrain slice \\
\midrule
\multicolumn{3}{@{}l}{\emph{Serving (Section~\ref{sec:method:deploy})}} \\
Cache & in-memory KV & Keyed by $(q,d_{\text{last,id}})$ \\
Coverage & ${>}90\%$ & Rewriter-impacted reqs \\
\midrule
\multicolumn{3}{@{}l}{\emph{A/B evaluation (Section~\ref{sec:exp:online})}} \\
Traffic/group & $10$--$30\%$ & Randomly assigned users \\
\bottomrule
\end{tabular}
\end{table}

\section{Algorithm: Semi-Online MPO Reward Finetuning}
\label{app:algorithm}

\begin{algorithm}[H]
\caption{Semi-Online MPO Reward Finetuning}
\label{alg:semionline}
\begin{algorithmic}[1]
\Require Initial policy $\pi_{\theta_0}$, training set $\mathcal{T}$, reward $r$, phase count $P$
\For{$p = 1, \dots, P$}
  \State $\mathcal{R}_p \gets \emptyset$
  \For{each $(q, d_{\text{last}}, \mathcal{D}^+, \mathcal{D}^-) \in \mathcal{T}$}
    \State Sample $\{\tilde{q}^{(1)}, \dots, \tilde{q}^{(N)}\} \sim \pi_{\theta_{p-1}}(\cdot \mid q, d_{\text{last}})$
    \State Compute $r(\tilde{q}^{(m)})$ for $m = 1, \dots, N$
    \State Construct preference pairs $\mathrm{Pairs}(\cdot) \subseteq \{(\tilde{q}^+, \tilde{q}^-)\}$ \Comment{top-$k$ $\times$ bottom-$k$ with margin filter and raw-query branch, see \S\ref{sec:method:reward}}
    \State $\mathcal{R}_p \gets \mathcal{R}_p \cup \{(q, d_{\text{last}}, \tilde{q}^+, \tilde{q}^-) : (\tilde{q}^+, \tilde{q}^-) \in \mathrm{Pairs}(\cdot)\}$
  \EndFor
  \State $\theta_p \gets \mathrm{MPOEpoch}(\theta_{p-1}, \mathcal{R}_p, \eta_p)$ \Comment{cosine restart at phase boundary}
\EndFor
\State \Return $\pi_{\theta_P}$
\end{algorithmic}
\end{algorithm}

\end{document}